\documentclass[manuscript]{aastex}
\usepackage{epsfig}

\usepackage{amssymb}
\newcommand{\kms}{~km~s$^{-1}$}
\shorttitle{WD mass in GW Lib}
\shortauthors{van Spaandonk et al.}

\begin{document}
%last edit:
\title{The mass of the white dwarf in GW Libra}
\author{L. van Spaandonk}
\affil{Astronomy and Astrophysics, Dept. of Physics, University of Warwick, Coventry CV4 7AL, UK}
\email{l.van-spaandonk@warwick.ac.uk}

\author{D. Steeghs}
\affil{Astronomy and Astrophysics, Dept. of Physics, University of Warwick, Coventry CV4 7AL, UK}
\affil{Harvard-Smithsonian  for Astrophysics, 60 Garden Street, Cambridge, MA 02138, USA}

\author{T.~R. Marsh, S.~G. Parsons}
\affil{Astronomy and Astrophysics, Dept. of Physics, University of Warwick, Coventry CV4 7AL, UK}

%\begin{abstract}
%We have measured the gravitational redshift of the pulsating white dwarf component of the  WZ~Sge type Dwarf Nova GW~Lib from archival VLT/UVES data. This, combined with empirical mass radius models for WDs, provides us with a direct measurement of the WD mass of $M_1 = 0.84 \pm 0.02 M_\odot$. The \textsc{fwhm} gives a $v \sin i = 86.97 \pm 3.35$\kms, which results in a spin period of the WD of $97 \pm 12$~s.   
%\end{abstract}
\begin{abstract} We report a mass and rotational broadening ($v\sin i$) for the pulsating 
white dwarf component of the WZ~Sge type Dwarf Nova GW~Lib based on 
high-resolution VLT spectroscopy that resolves the Mg~\textsc{ii} 
4481\AA~absorption feature. Its gravitational redshift combined with 
white dwarf mass-radius models, provides us with a direct measurement 
of the white dwarf mass of $M_1 = 0.84 \pm 0.02 M_\odot$. The line is 
clearly resolved and if associated with rotational broadening gives
$v \sin i = 87.0 \pm 3.4$\kms, equivalent to a spin period of $97 
\pm 12$~s.
\end{abstract}
\keywords{binaries: spectroscopic --- novae, cataclysmic variables --- stars: individual (GW Lib)}

\section{Introduction}

The population of cataclysmic variables (CVs) provides an important sample of binary systems possessing homogeneous configurations of white dwarfs (WDs) accreting from near main-sequence donor stars.  A CV initially evolves towards shorter orbital periods, but near a period of $\sim 75$min bounces back towards longer periods with its donor star turning into a degenerate brown dwarf. Recent CV searches, such as the significant harvest of CVs from SDSS \citep{szkody09-1, gaensicke09-1} are finally unearthing large numbers of systems at short orbital periods as expected from binary evolution considerations.

Reliable estimates for binary parameters are needed to place individual systems on their evolutionary tracks.
The faintness of the low mass donor stars in comparison to the WD and its accretion flow makes such parameter estimates difficult for short period systems (e.g WZ Sge; \citealt{steeghsetal01-2}; \citealt{steeghsetal07-1}), except in favourable cases where eclipse constraints can be exploited \citep{littlefairetal06-2}. 
\citet{patterson01-1} has identified an indirect method of inferring binary mass ratios using the tidally driven superhump modulations in the lightcurves of CVs. Although promising, this furthermore highlights the need for suitable calibrator systems for which binary parameters can be accurately determined independently.

GW Lib is a short period dwarf nova ($P_{\mathrm{orb}}$=76.78 mins; \citealt{thorstensenetal02-3}) displaying the typical characteristics of a CV near its period minimum. 
Crucially, it was the first CV discovered to contain a pulsating WD \citep{vanzyletal04-1}, by analogy with field WD pulsators that lie in the instability strip for non-radial pulsations. 
So far, various methods to determine its system parameters have been exploited, including model fits to the WD absorption profiles, prominent in both the optical \citep{thorstensenetal02-3} and the UV \citep{szkodyetal02-4} providing estimates for the WD $T_{\mathrm{eff}}$ and $\log g$. Asteroseismological models, combined with a UV-flux limit, suggests a WD mass of $M_1 = 1.02 M_\odot$ (\citealt{townsleyetal04-1}). After its second recorded super-outburst in April 2007 \citep{templeton07-1}, the superhump period suggest a mass ratio of $q=M_2/M_1=0.06$ \citep{katoetal08-1} when combined with Patterson's empirical relation \citep{pattersonetal05-3}. 
%Finally, several classical dynamical constraints have been published using the various emission lines present in GW~Lib's quiescent spectrum.  Pre-outburst Balmer lines have been used to determine the orbital period \citep{thorstensenetal02-3} while post-outburst time-resolved spectroscopy revealed donor emission features in the Ca~\textsc{ii} lines \citep{vanspaandonketal09-1}. 

In this letter we explore the possibilities of using the gravitational redshift of the Mg~\textsc{ii} absorption line reported in \citet{vanspaandonketal09-1} in combination with mass-radius relations to give an independent measurement of the   mass and spin of the WD in GW Lib. 

\section[]{Observations and reduction}
\label{sec:observations}

GW~Lib was observed within the 69.D-0591 program at the 8.4-m Very Large Telescope (\textsc{vlt}), located at the Paranal Observatory in Chile as part of the European Southern Observatory, equipped with only the blue arm of the Ultraviolet and Visual Echelle Spectrograph (\textsc{uves}: \citealt{dekkeretal00-1}).  We retrieved a total of 45 science frames obtained during 2002 May 16-17 covering 2.4 and 2.8 binary orbits on the respective nights. 
The reduction of the raw frames was conducted using the most recent standard recipe pipeline release of the \textsc{uves} Common Pipeline Library (\textsc{cpl}) recipes. The resultant optimally-extracted spectra covered a wavelength range of $\lambda$4020-5240\AA\ at a dispersion of 0.031\AA~pixel$^{-1}$ and a spectral resolution of 0.10\AA~(5.77\kms) as measured from the skylines.
The spectra were wavelength calibrated with one Thorium Argon arc per night. This calibration was tested against the two sky lines visible at $\lambda$5197.92\AA\, and $\lambda$5200.28\AA. The corresponding science frames were corrected for any remaining shifts. The residuals were scattered around zero with a maximum amplitude of $0.01$~km~s$^{-1}$. Next, heliocentric velocity corrections were applied to the individual frames to deliver spectra in a common heliocentric rest frame.

\begin{figure}
  \figurenum{1}
  \epsscale{1.0}
  \plotone{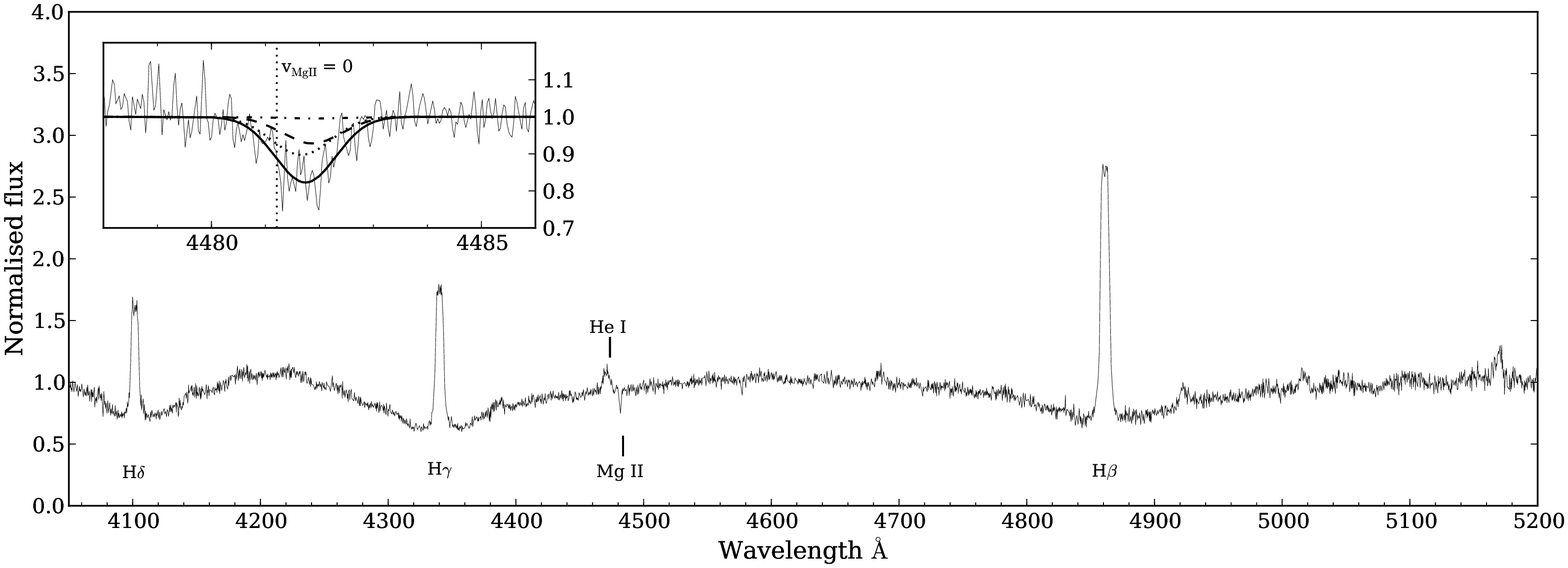}
  \caption{Average spectrum of GW Lib on the 16th of May 2002. The inset shows the Mg~\textsc{ii} absorption line averaged over 1 orbital period with the final Gaussian fit. The total fit (\textit{solid line}) is the sum of three separate fits to the triplet lines, all according to their own transition probability. $\lambda$4481.126 \textit{dotted}, $\lambda$4481.150 \textit{dotted-dashed}, $\lambda$4481.325 \textit{dashed}.}
  \label{fig:spectrum}
\end{figure}

No standard was observed in the correct settings on the nights and as no master response curve exists for the non-standard setup used, the frames were not flux calibrated. The exposure time was 500 seconds giving a signal to noise of $\sim$6.5 per spectrum.  Details of the observations can be found in Table \ref{tab:observations} while Figure \ref{fig:spectrum} shows the average spectrum of GW~Lib on the 16th of May. Prominent features are the Balmer disc emission lines on top of broad absorption troughs from the WD, visible due to the low mass accretion rate in the system.  He~\textsc{i} is also seen in emission as is He~\textsc{ii} at 4685.75\AA. 

Our pre-outburst, intermediate resolution spectra of GW Lib showed Mg~\textsc{ii} at 4481.21\AA\ in absorption (see figure 2, \citealt{vanspaandonketal09-1}). 
The archival \textsc{vlt/uves} data confirm the presence of this line and thanks to the superior spectral resolution combined with the low inclination of the system shows it to be unblended from the nearby He~\textsc{i} emission at 4471\AA. 
We measure an EW of $0.25\pm0.01$\AA\ (similar to $0.24\pm 0.03$\AA\, in 2004) and a Full Width at Half Maximum (\textsc{fwhm}) of $1.35\pm 0.04$\AA.

\section[]{Gravitational Redshift for Mg II}
\label{sec:massdetermination}
In low accretion rate dwarf novae, the luminosity of the accretion disc is low enough to show the broad absorption features of the WD and can even show the narrow absorption features of metal lines due to freshly accreted gas. These lines open a window to probe the WD atmosphere directly and give independent measurements of stellar parameters.
%(\citealt{sionetal94-1}: \citealt{sionetal97-1}; \citealt{smithetal06-1}; \citealt{steeghsetal07-1}). 
For lines formed near the primary, a gravitational red-shift is expected, introduced in the deep gravitational potential of the WD (\citealt{eddington24-1}; \citealt{greenstein+timble67-1}; \citealt{sionetal94-1}). A measurement of the gravitational redshift in the rest frame of the binary could provide the WD mass directly when combined with mass-radius models (e.g. Eggleton's relation as quoted in \citealt{verbunt+rappaport88-1}). This method has previously been used in CVs in the cases of U~Gem \citep{long+gilliland99-1}, VW Hyi \citep{smithetal06-1} and WZ~Sge \citep{steeghsetal07-1}.

In the case of GW Lib,  the directly measured redshift of the magnesium line ($v_{\mathrm{MgII}}$) needs to be rectified for several contributions in order to get the true gravitational redshift induced by the WD only ($v_{\mathrm{grav}}(\mathrm{WD})$). These corrections consist of the systemic velocity ($\gamma$) of the binary system and the effects of the gravitational potential from the donor star ($v_{\mathrm{grav}}(\mathrm{donor})$). Hence the gravitational redshift due to the WD is given by:

\[
v_{\mathrm{grav}}(\mathrm{WD}) = v_{\mathrm{MgII}} - v_{\mathrm{grav}}(\mathrm{donor}) -\gamma 
\]

We will discuss the various contributions and measurements independently. Firstly, all 45 spectra of the two nights were individually continuum normalised and then binned into 20 equally spaced orbital phase bins to increase S/N. 

\subsection{$v_{\mathrm{MgII}}$}
To measure the redshift of the magnesium line to the best precision and minimise any orbital effect we compared several methods.

Firstly, we made a weighted orbital average and fitted the Mg~II absorption line with a weighted triple Gaussian as the Mg~\textsc{ii} is a triplet line. The different components have rest wavelengths of $\lambda$4481.126\AA\, with a  transition probability of $\log(gf)= 0.7367$,  $\lambda$4481.150\AA\, with $\log (gf) = -0.5643$  and $\lambda$4481.325\AA\, with $\log (gf) = 0.5818$\footnote{Data from The Atomic Line List Version 2.04: \texttt{http://www.pa.uky.edu/\~ {}peter/atomic/} }. The variables for the fit are the common offset (for all 3 lines) and the common \textsc{fwhm} to provide a good fit to the blue wing of the absorption feature. The peak height is also a common variable but scaled according to the various transition probabilities. This gives a best fit with a mean offset of $35.2 \pm 1.1$\kms, a \textsc{fwhm} of $1.32\pm 0.08$\AA\, and a common absorption line depth of $2\pm0.1$\%. 

Secondly, we fixed the \textsc{fwhm} and peak of the Gaussians to these values and checked the individual spectra for orbital variablity. Unfortunately, the S/N is too low to give good individual fits (individual $1\sigma$ errors on the offset are $\sim 8$\kms)  nor can we phase-lock this motion to the ephemeris of the system as determined in \citet{vanspaandonketal09-1} due to the uncertainty in the period. 
The resultant radial velocity curve suggested motion, and a formal sine fit delivered a semi-amplitude of $K = 13 \pm 2 $\kms. To rectify for any orbital motion, we removed any measured shift compared to the mean from individual spectra.

An uniform orbital average was constructed to minimise any effects caused by varying S/N over the orbital period and finally we refit this last average with our triple Gaussian fit with as variables the common offset, the common \textsc{fwhm} and the weighted peak as described before (see inset Figure \ref{fig:spectrum}).

This gives a final measurement of $v_{\mathrm{MgII}} = 35.8 \pm 1.5$\kms. The line has a  \textsc{fwhm} $= 1.30 \pm 0.05$\AA. We have tested the accuracy of our measurement by following different recipes for combining and averaging the spectra but note that the uncertainty on the gravitational redshift is dominated by the S/N and resolution of the data and not the specific recipe used.

\subsection{$ v_{\mathrm{grav}}(\mathrm{donor})$}
The first correction is due to the influence of the gravitational potential of the donor on the magnesium line. However, the expected low mass of the donor star ($\sim 0.05~M_\odot$) gives only a small correction of $0.06\pm 0.02$\kms\, near the WD surface, effectively negligible given the measurement uncertainty.

\subsection{The systemic velocity $\gamma$}
\label{subsec:gamma}
From simultaneous double Gaussian fits to the double peaked H$\beta$ and H$\gamma$ disc lines, combined with radial velocity curve fits provide a systemic velocity of $-12.3 \pm 1.2$\kms\, and a semi-amplitude of $K_{\mathrm{disc}} = 36.4\pm 1.8$\kms.  These values are consistent with previously derived values \citep{thorstensenetal02-3,vanspaandonketal09-1}.

As we measure $\gamma$ from disc lines, we need to take into account that this  emission is red-shifted both by the WD and the donor star. As $v_{\mathrm{grav}}$ is reciprocal to the distance, the amount we have to account for is minimal at the edge of the disc, $R_{\mathrm{outer\,disc}} = 2.2\times 10^{8}$~m (Equation 2.61 of \citealt{warner95-1}) giving a minimal correction to $v_{\mathrm{grav}}$ of $1.7$\kms. A more realistic value comes from the projected Keplerian velocity at the edge of the disc $K_{\mathrm{disc}} \sim 200$\kms\, (from the location of the disc ring in Doppler maps and the estimate for $i$, see \citealt{vanspaandonketal09-1}) which gives a Kepler speed of $v_{\mathrm{disc}} = K_{\mathrm{disc}}/ \sin i  \sim 1070$~km~s$^{-1}$ and corresponds to a $v_{\mathrm{grav}} = v^2/c \sim 3.8$\kms. Including the transverse Doppler redshift (at half the strenght of the gravitational redshift) this amounts to a total gravitational redshift at the location of the disc lines of $5.7\pm 1.6$\kms. The gravitational potential of the donor star has an effect of only $0.08\pm0.02$\kms\, at a distance of $\sim 3 \times 10^{8}$~m. Hence the total $v_{\mathrm{grav}}\mathrm{(disc)}=5.8 \pm  1.6$\kms. 

The $\gamma$ from the disc lines combined with the above correction should be consistent with the $\gamma$ suggested by the radial velocity curve from the donor star (corrected for the effects of the gravitational potential at its surface).
From the Ca~\textsc{ii} emission line in the \textsc{i}-band we previously found $\gamma = -13.1 \pm 1.2$ \citep{vanspaandonketal09-1}, the correction at the surface of the donor star is $0.34\pm 0.15 $\kms\, from the donor star and $1.4\pm0.2$\kms\, from the WD (again including the transverse Doppler shift). 
Thus we have $\gamma_{\mathrm{disc}}= (-12.3 \pm 1.2)- ( 5.8\pm 1.6) = -18.1 \pm 2.0 $ and $\gamma_{\mathrm{donor}}= (-13.1 \pm 1.2)- (1.7 \pm 0.3) = -14.8\pm 1.2$. Both estimates for $\gamma$ are indeed consistent and with similar precision. Hence we have used $\gamma_{\mathrm{disc}}$ as it is derived from the same data set.

\subsection{The implied WD mass}
Combining all values we find for the final gravitational redshift:
\begin{eqnarray}
v_{\mathrm{grav}}\mathrm{(WD)}&=& (35.8 \pm 1.5) - (0.06 \pm 0.02)  -(-18.1 \pm 2.0) \, \mathrm{km\,s}^{-1} \nonumber\\
 &=& 53.8 \pm 2.5\, \mathrm{km\,s}^{-1}\nonumber  
\end{eqnarray}
When combined with theoretical and empirical models for the mass-radius relationship for WDs this gives a direct measurement of the WD mass. The models we used are Eggletons zero-temperature mass-radius relation as quoted by \citet{verbunt+rappaport88-1} and several appropriate non-zero temperature models for GW Lib from \citet{fontaineetal01-1}. 
We plot these models in Figure \ref{fig:mass_radius_relation} together with the $M_1(R)$ line demanded by our measured $v_{\mathrm{grav}}$. Accommodating the intersections between non-zero temperature models we find that $M_1 = 0.84\pm0.02~ M_\odot$. 

\subsection{System parameters and WD spin}

\citet{katoetal08-1} reported the detection of superhump modulations in GW Lib. Combining this with the period from \citet{thorstensenetal02-3} and the improved superhump excess - mass ratio relation given by \citet{knigge06-1}, the implied mass ratio of the system is $q = 0.060 \pm 0.008$.  We previously determined the projected radial velocity of the donor star, $K_2$ in \citet{vanspaandonketal09-1}. Thus with an independent determination of $M_1$ from this paper, we can solve the system parameters for GW Lib under these constraints  and list these in Table \ref{tab:systemparameters}.

If one assumes that the width of the Mg~\textsc{ii} absorption line is dominated by rotational broadening, its \textsc{fwhm} can constrain the spin period of the WD.  The \textsc{fwhm} of the absorption line fit is measured to be $1.30 \pm 0.05$\AA\, which corresponds to a  $v \sin i $ of $87.0\pm 3.4$\kms\, at this wavelength. Assuming $i$ is close to the value giving in Table \ref{tab:systemparameters}, this translates into a rotation speed of $448\pm 24$\kms\, at the surface of the WD. For a WD of mass $0.84\pm 0.02~M_\odot$ and a radius for $6.95\pm 0.15\times 10^{8}$~cm this results in a spin period of the WD of $97 \pm 12$~seconds.

\section{Discussion}
\label{sec:discussion}
We have measured the gravitational redshift of the Mg~\textsc{ii} absorption line in high-resolution echelle spectra of GW~Lib during quiescence. Assuming an origin in the photosphere of the accreting WD, we combined this redshift with non-zero temperature mass-radius relations for WDs to derive a WD mass of $0.84\pm 0.02M_\odot$. Combining this independent measurement with other constrains confirms that GW~Lib is a low mass ratio system observed at very low inclination (Table \ref{tab:systemparameters}). Because the Mg~\textsc{ii} line was well-resolved in our data, we could also estimate the spin period of the WD to be $97 \pm 12$~seconds if we assume the width of the line to be dominated by rotational broadening.

\begin{figure}
  \figurenum{2}
  \epsscale{0.4}
   \plotone{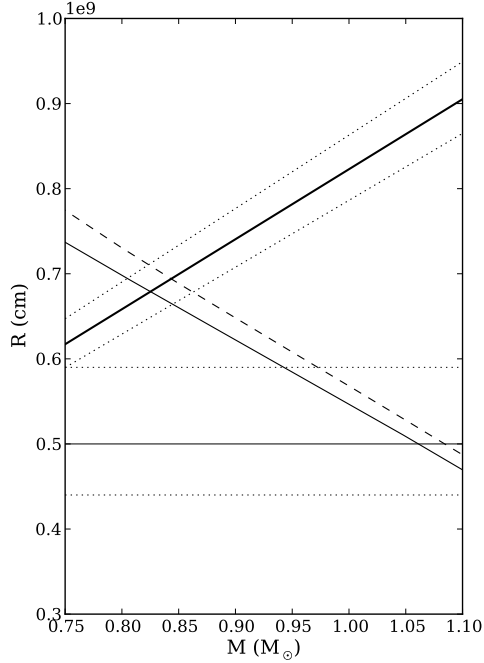}
   \caption{Mass radius relation ship for the measured $v_{\mathrm{grav}}=53.8$~km~s$^{-1}$ (\textit{dashed line from lower left to upper right}) with 1$\sigma$ error lines (\textit{dotted lines}). The mass-radius models are Eggleton's zero-T relation (\textit{solid line from upper left to lower right}),  a $T=15\,500$~K, $q\mathrm{H}=10^{-4}$ and $q\mathrm{He} = 10^{-2}$ C-core model (\textit{dashed line}). Models with the same core but diferent temperatures at $T=18\,000$~K and $T=13\,000$~K are similar and hence not plotted.  For a WD with a CO-core, $T=15\,500$~K, $q\mathrm{H}=10^{-10}$ and $q\mathrm{He} = 10^{-2}$, the model is almost similar to the zero-T relation. The horizontal lines denote the UV flux estimate from the Szkodys paper corrected for the latest distance measurement for GW Lib (\textit{solid line}) and  $1\sigma$ errors (\textit{dotted})}
  \label{fig:mass_radius_relation}
\end{figure}

Our mass value is significantly below the $1.02~M_\odot$ lower limit derived by \citet{townsleyetal04-1}. However, their choice of preferred solution was largely driven by the WD size as implied by the measured UV flux. 
\citet{szkodyetal02-4} fitted WD models to UV spectra of GW Lib. Two different single temperature solutions were explored;  $d=171$~pc ($M_1 = 0.6 M_\odot / v_{\mathrm{grav}} \sim 29$\kms) and $d=148$~pc ($M_1 = 0.8 M_\odot / v_{\mathrm{grav}} \sim 49$\kms) respectively. These solutions can be corrected to the latest parallax distance of $100^{+17}_{-13}$~pc (Thorstensen, private communication - improvement of the distance in \citealt{thorstensenetal02-3}). At this distance, the observed UV flux implies a radius of $5\times 10^{8}$~cm (for a WD temperature of $14\,700$~K). However, such a small radius intercepts the mass-radius relations at $1.07~M_\odot$, Figure \ref{fig:mass_radius_relation} (\textit{horizontal solid line}) and was indeed the reason \citet{townsleyetal04-1} ruled out their lower mass solutions. Such a massive WD would result in $v_{\mathrm{grav}} \sim 95$\kms, much larger than our measured redshift would suggest. 
%Our gravitational redshift would only support such a massive WD if the Mg~\textsc{ii} absorption arises from a height that is 1.4 times the WD radius, which seems rather unlikely.
% 
We note that \citet{szkodyetal02-4} fail to find a good single temperature fit and claim the best fit arises fitting a dual temperature model to the spectrum, allowing for a cooler zone that would place GW Lib within the ZZ Ceti instability strip for pulsating single WDs. However, the UV-flux versus implied WD radius then becomes a less clear-cut argument given the freedom of two temperatures and the surface coverage split between the two.
We note that the reverse has been seen in U~Gem where the UV-flux places the WD at a lower mass than the gravitational redshift suggests \citep{long+gilliland99-1, longetal06-1}. 
%Additional UV spectroscopy and model fitting is warranted to try and achieve a picture consistent with the latest distance and our dynamical constraints from optical spectroscopy.

As \citet{townsleyetal04-1} warn, no WD rotation was included in their asteroseismology models and therefore their derived parameters ($M_1, M_{\mathrm{acc}}, \dot{M}$) may need to be revisited given our estimate of the spin period of the WD in GW Lib. A  $P_{spin}= 97 \pm 12$~seconds should give rise to rotationally split modes. \citet{vanzyletal04-1} conducted a thorough campaign with a baseline of over 4 years to find these modes in the power spectrum of GW Lib. The extensive search revealed a possible doublet around the 230s mode with a frequency difference between the components of $0.79$~$\mu$Hz. If this doublet originates from the rotation splitting of an $l=1$ mode, the WD would have a spin period of $\sim 7.3$~days which is much longer than expected for the WD in a CV.  Approached from the other side, if a similar mode is split as a result of a rotation period of $\sim 100$~s, the frequency difference would be $\sim 0.5$~mHz and should be visible in the power spectrum. Note that for the proposed spin period, the first order approximation for frequency splitting is no longer valid and the splitting would become asymmetric.  
%For conclusive models more than 3 modes are needed (especially as they identify the 3 frequencies as $n=$3, 8 and 17 modes). 

From both GW~Lib's and U~Gem's comparison between WD mass found by gravitational redshift and the constraints from the UV-flux we can conclude that the measurement of the gravitational redshift through WD absorption lines gives in principle a good measurement of the WD mass in CVs, but is not always fully consistent with other constraints. Nonetheless, it suggests photospheric absorption lines in the atmospheres of accreting WDs can be a viable tool for mass measurements, in particular for low mass accretion rate CVs where the light from the WD dominates and especially for low $i$ systems.

For GW Lib itself, we are still dependent on a number of assumptions in order to solve its system parameters. To properly align all methods, we need to revisit the mass from asteroseismology taking into account the effects of rotation and see if a WD model with a mass of $0.84 M_\odot$ at the latest distance can be fitted to the UV data with reasonable temperatures. To be able to calculate a direct mass ratio, a determination of $K_1$ from the Mg~\textsc{ii} should be possible through high S/N phase-resolved spectroscopy. If furthermore matched with an solid ephemeris using the emission from the donor star (giving $K_2$ and $\gamma$), a fully consistent and accurate WD mass for this prototypical accreting WD pulsator is entirely feasible.

\acknowledgments
DS acknowledges an STFC Advanced Fellowship. TRM was supported by an STFC Rolling Grant. Based on observations made with ESO Telescopes at the Paranal Observatory under programme ID 69.D-0591

\bibliographystyle{mn_new}                                              
%\bibliography{my_bib,aabib,proceedings}

\begin{table}
  \begin{center}
    \caption{Details of observations}
    \label{tab:observations}
    \begin{tabular}{l l l l l}
      \hline
      \hline
      Date & $n$ spectra& Exp time & Start/Finish  & Orbital Coverage\\
      &                     & (s)  & UTC & ($T_{\mathrm{obs}}/P_{\mathrm{orb}}$)\\
      \hline
      16 May 2002 & 20 & 500 & 05:28:26 - 08:34:02 & 2.4\\
      17 May 2002 & 25 & 500 & 04:49:03 - 08:24:08 & 2.8\\
      \hline
    \end{tabular}
  \end{center}
\end{table}

\begin{table}
  \begin{center}
    \caption{System parameters for GW Lib}
    \label{tab:systemparameters}
    \begin{tabular}{ l l l }
      \hline
      \hline
      $P\,(\mathrm{min})$ & 76.78$\pm$ 0.03 & \citealt{thorstensenetal02-3} \\
      $M_1\,(M_\odot)$ & $0.84 \pm 0.02$  & This paper\\
      $q$ & $0.060\pm 0.008$  & \citealt{katoetal08-1, knigge06-1}\\
      $K_2\,( \mathrm{km \, s^{-1}})$ & $100.8 \pm 7.1$ & \citealt{vanspaandonketal09-1}\\
      $K_1\,( \mathrm{km \, s^{-1}})$ & $6.25 \pm 0.4$ & From $q$ and $K_2$\\
      $M_2\,(M_\odot)$ & $0.05 0 \pm 0.007$  & From $q$ and $M_1$\\
      $v_1\,( \mathrm{km \, s^{-1}})$ & $30.8 \pm 0.5$ & Kepler 3rd Law\\
      $v_2\,( \mathrm{km \, s^{-1}})$ & $513.4 \pm 4.8$ &  Kepler 3rd Law \\
      $i\, (^{\circ}) $ & $11.2 \pm 0.4$ & From $K_2$ and $v_2$ \\
      $P_{\mathrm{spin}} (\mathrm{s})$ & $97 \pm 12$ & From $v \sin i$ \\ 

      \hline
    \end{tabular}
  \end{center}
\end{table}

\end{document}